# Polarization of ultrashort optical pulses in semiconductor superlattice in a presence of magnetic field

M.B. Belonenko[1], E.N. Nelidina[2]

[1] *Nanotechnology laboratory, Volgograd Business Institute,*

*Yujno-Ukrainskaya st. 2, 400048, Volgograd, Russia*

[2] *Volgograd State Medical University,*

*Pl. Pavshikh Bortsov Square 1, 400131, Volgograd, Russia*

The system of equations describing an ultrashort optical pulses propagation in semiconductor superlattice with applied magnetic field was obtained based on Boltzmann equation in the relaxation-time approximation for the single electron distribution function and coupled Maxwell equations for the electromagnetic field. It was demonstrated that an original linear-polarized optical pulse initiates an orthogonal polarization in the field of a sample. The propagation dynamics of initiated and initial pulses in the sample was investigated. The effects of the lattice geometry have been elicited.

1. **Introduction**

Modern nanotechnologies allow creating systems of complex geometry. This stimulates theorists' interest in investigation of objects such as quantum wires, rings, cylinders and wells. The peculiarities of these objects geometry affect on current carriers' spectrum, consequently determine specific features of electronic properties of such structures. Thus, a semiconductor superlattice represented as a structure in which the electron is effected both by a crystal lattice potential and additional simulated potential with a constant greater than that of a crystal lattice is a material in which nonlinear effects can be observed even in moderately intense alternating fields. The additional potential leads to division of crystal energy bands close to their edges; owing to that fact essential nonlinearity of superlattice electronic properties shows itself in moderately intensity fields [1,2].

Apparently the easiest way to construct a superlattice is connected with producing a regular system of quantum wells in a solid body. These quantum wells bound by the tunnel effect give the dispersion law of current carriers. This approach let us not only set the value and dispersion law for minizone but also produce superlattices with decreased dimension just changing the distance between quantum wells. Interest in such structures has grown owing to Tanamoto's model of quantum computer [3,4].

It is very important that electromagnetic waves propagating in superlattice structures become highly nonlinear even at relatively weak fields (2-3 orders weaker than in standard semiconductor materials). One of the consequences is the possibility of solitary electromagnetic wave propagation in superlattices, which are the analogs of solitons or even solitons. Superlattices properties under consideration initiated both an intense academic interest and attempts to be applied in nonlinear optics devices. The investigation of propagation of ultrashort light pulses (optical solitons) in the superlattices can be mentioned as an instance [5, 6]. At the same time, it should be noted that electronic properties which can be displayed in the optical part of the spectrum are often out of consideration. Particularly, there can be the situation when keeping the spectrum of the current carrier under control, for instance, by means of applied magnetic field we can manipulate the important optical pulses characteristics which include polarization. From physical standpoint this is caused by phenomena similar to Hall effect, that is the deflection of moving under electromagnetic field pulse electrons by an external magnetic field.

In this context the research problem of the ultrashort optical pulses propagation in superlattices in the case of applied external magnetic field which can change a pulse polarization mode is arisen. High-power ultrashort electromagnetic pulses when it is possible to neglect the relaxation processes during the pulse action are of a special interest. This involves a pulse propagation in mode similar to self-induced transparency one and thus the pulse energy loss in a sample is not taking place.

Summing all the mentioned above it may be concluded that the research of nonlinear electromagnetic waves dynamics and investigation of their polarization in superlattices with applied external magnetic field is a critically important and urgent problem.

## 2. Basic equations

Let us consider two-dimensional superlattice with spectrum given by the Fourier series:

$$\varepsilon(p_x, p_y) = -\Delta \sum_{m,n=0} A_{mn} \cos map_x \cos nap_y \tag{1}$$

where $\Delta$ is a minizone width, $a$ is a superlattice constant, $p_x, p_y$ are electron quasi-momentum components, posed in the first Brillouin zone. Supposing that an ultrashort optical pulse propagates in the direction of z-axis (perpendicular to the superlattice plane) we shall describe the pulse electromagnetic field classically, in terms of Maxwell equations. Thus, in case under investigation, in the gauge $\vec{E} = -\frac{1}{c}\frac{\partial \vec{A}}{\partial t}$, Maxwell equations with regard to the dielectric and magnetic properties of semiconductor superlattice [7] can be written in the form:

$$\frac{\partial^2 \vec{A}}{\partial z^2} - \frac{1}{c^2}\frac{\partial^2 \vec{A}}{\partial t^2} + \frac{4\pi}{c}\vec{j} = 0 \tag{2}$$

here we disregard the diffraction spreading of the laser beam in the directions perpendicular to the axis of the pulse propagation. Current density has the form $\vec{j} = (j_x(z,t), j_y(z,t), 0)$, and it is assumed that the vector potential $\vec{A}$ has the form $\vec{A} = (A_x(z,t), A_y(z,t), 0)$.

The electric current is determined with semiclassical approximation [8], with dispersion law (2) taken from a quantum-mechanical model, and the evolution of an ensemble of particles described by the Boltzmann kinetic equation in the relaxation-time approximation, that is

$$\frac{\partial f}{\partial t} + (-\frac{q}{c}\frac{\partial A_x}{\partial t} + qhv_y)\frac{\partial f}{\partial p_x} + (-\frac{q}{c}\frac{\partial A_y}{\partial t} - qhv_x)\frac{\partial f}{\partial p_y} = \frac{F_0 - f}{\tau} \quad (3)$$

where $h$ is magnetic field parallel to z-axis applied to the sample, and $v_x = \partial \varepsilon / \partial p_x$, $v_y = \partial \varepsilon / \partial p_y$, the equilibrium function $F_0$ of the Fermi distribution can be written in the form $F_0 = \frac{1}{1 + exp(E(\vec{p})/k_b T)}$, $T$ is a temperature, $k_b$ is the Boltzmann's constant. Since the typical pulse duration of ultrashort pulse is about $10^{-15}$ s and the relaxation time $\tau$ is estimated as $10^{-12}$ s according to [9], the equation (3) can be solved in the collisionless limit $\tau \to \infty$. Let us use the following expression for the current component $j_\alpha(z,t)$ ($\alpha = x, y$):

$$j_\alpha = q \int dp v_\alpha f \quad (4)$$

Then let us use the method of an average electron [10-12], wherein the current is expressible as a solution of the classical motion equation for the electron in assumed fields:

$$\frac{dp_x}{dt} = -\frac{q}{c}\frac{\partial A_x}{\partial t} + qhv_y$$
$$\frac{dp_y}{dt} = -\frac{q}{c}\frac{\partial A_y}{\partial t} - qhv_x \quad (5)$$

with initial conditions $p_x|_{t=0} = p_{x0}$; $p_y|_{t=0} = p_{y0}$. For simplicity in the instant case of low temperatures and in the limit $\tau \to \infty$ this method also gives an expression:

$$j_\alpha = qn_0 v_\alpha(p_x, p_y) \quad (6)$$

where $p_x, p_y$ are the solutions of (5) with initial conditions $p_{x0} = p_{y0} = 0$. It should be noted that in the case of high temperatures in the expression (6) it is necessary to use the solutions of the equations (5) with arbitrary initial conditions, and then strike an average with the equilibrium distribution function wherein the initial conditions for $p_x, p_y$ will act as pulses.

In general case, it is difficult to obtain a solution for the whole range of alternating electric and constant magnetic fields for the dispersion law given by (1). For instance, the solutions described in [10,13] for the cosine dispersion law and for the parabolic dispersion law are known. Let us use the fact that the amplitude of ultrashort optical pulses is sufficiently high and it is possible to seek a power series solution in powers of $h$. Then with basic terms we can obtain:

$$p_x = -\frac{q}{c} A_x + qh \int_0^t dt \frac{\partial \varepsilon(-\frac{q}{c} A_x, -\frac{q}{c} A_y)}{\partial(-\frac{q}{c} A_y)}$$

$$p_y = -\frac{q}{c} A_y - qh \int_0^t dt \frac{\partial \varepsilon(-\frac{q}{c} A_x, -\frac{q}{c} A_y)}{\partial(-\frac{q}{c} A_x)}$$

(7)

Thus, calculating the current according to (6) with pulses determined by (7) and plugging obtained value into (2) we get isolated self-consisted system of two equations in $A_x, A_y$. Solving this system it is possible to obtain a description of the field spatial distribution of ultrashort optical pulse and a description of its polarization.

These circumstances have given impetus to further numerical investigation of the equation (2,6,7), which has been obtained without any restrictions to the minimum duration of the electric field pulse.

### 3. The results of numerical analysis

The equations under investigation were numerically solved according to the cross-type finite-difference scheme [14]. The time and coordinate steps were determined from the standard conditions of stability. The steps of the finite-difference scheme were sequentially decreased by two until the solution changed in the eighth significant figure. The equations were non-dimensionalized in a standard way and further all the data will be represented in nominal units. For definiteness of the problem the dispersion law for the square lattice was chosen in the form:

$$\varepsilon(p_x, p_y) = \Delta(1 - \cos p_x - \sigma \cos 2p_x) + \Delta(1 - \cos p_y - \sigma \cos 2p_y)$$

(8)

where $\sigma < 1$ is a low parameter characterizing the spectrum with overrunning of the nearest neighbors taken into account [15,16]. The dispersion law for the superlattice consisting of regular triangular system of quantum wells was also chosen for a comparison:

$$\varepsilon(p_x, p_y) = \Delta(3 - \cos p_x - 2\cos(p_x/2)\cos(2p_y/\sqrt{3}))$$

(9)

An initial condition was chosen in the form of the well known kink-solution of the sine-Gordon equation:

$$A_x(z,0) = 4 \, arctg(\exp(z/\gamma))$$

$$\left.\frac{dA_x(z,t)}{dt}\right|_{t=0} = -\frac{2v}{\gamma} ch^{-1} z/\gamma$$

$$\gamma = (1-v^2)^{1/2} \qquad (10)$$

$$A_y(z,0) = \left.\frac{dA_y(z,t)}{dt}\right|_{t=0} = 0$$

This initial condition corresponds to the fact that the critical short pulse with plane polarization consisting of a single "half-oscillation" of the electric field is fed to the sample. An arising evolution of the electromagnetic field in time for the case of square lattice with the dispersion law (9) is shown in Fig. 1.

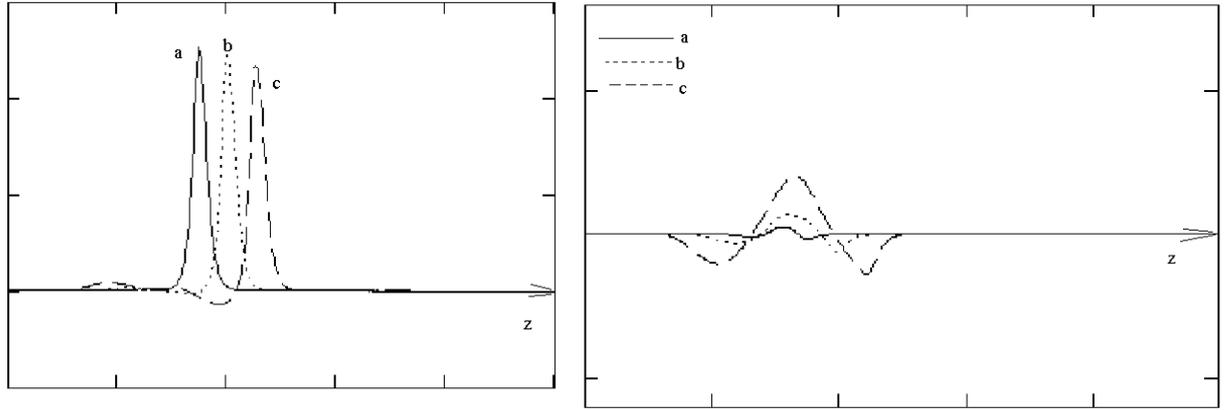

a) b)

Fig. 1. Electric field determined by the equations (2, 6, 7) at different instants of time. The non-dimensionalized coordinate (the unit is equal to $3 \bullet 10^{-4}$ m) is plotted along the *x*-axis, the non-dimensionalized value of electric field (the unit is equal to $10^7$ V/m) is plotted along the *y*-axis. Fig. a) the primary polarization field, fig. b) the orthogonal polarization field. The line b) indicates the time is 2 times greater than the line a), the line c) indicates the time is 3 times greater than the line a). v/c=0,95.

As it is seen from the presented result the separation of primary polarization pulse is realized simultaneously with the generation of orthogonal polarization pulse. Its amplitude grows in time, this is due to the current carrier velocity increase under the effect of the electric field, therefore more effective influence of constant magnetic field. It should be noted that the orthogonal polarization pulse has greater width, it seems, it is associated with the fact that the faster current carriers are accelerated more effectively therefore a spectrum is shrunk.

A form of ultrashort pulses of different polarizations varies depending on lattice geometry (Fig. 2).

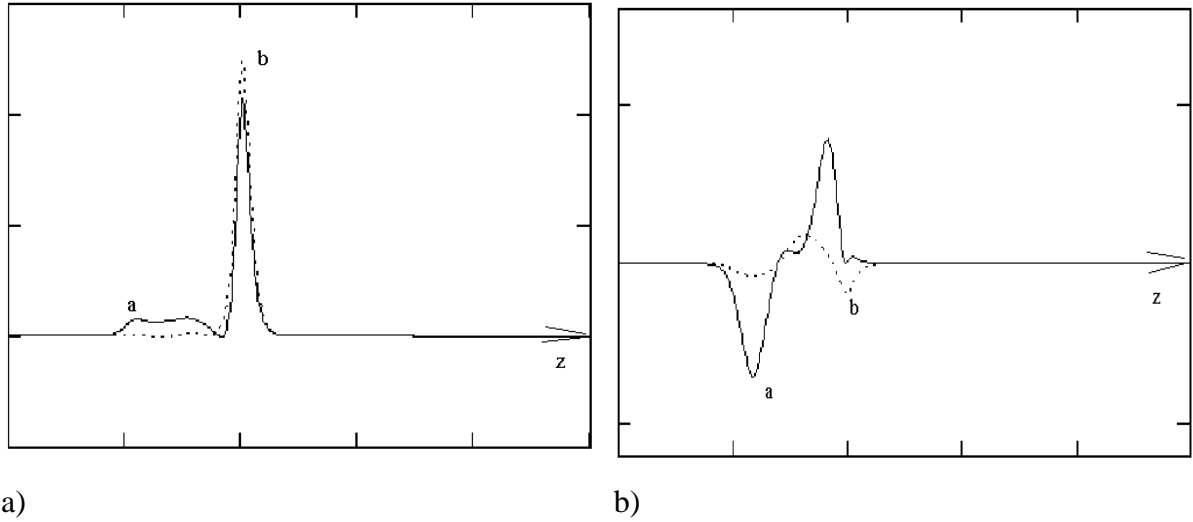

a)                                              b)

Fig. 2. Electric field determined by the equations (2, 6, 7) at different instants of time. The non-dimensionalized coordinate is plotted along the *x*-axis, the non-dimensionalized value of electric field (the unit is equal to $10^7$ V/m) is plotted along the *y*-axis. Fig. a) the primary polarization field, fig. b) the orthogonal polarization field. The line a) triangular lattice, the line b) square lattice. v/c=0,95.

In our opinion such geometry effect can be associated with the differences in the dispersion laws (8) and (9), namely that the dispersion law for the triangular lattice is non-additive and highly implicates current carriers pulses propagating in structure.

Dependence of ultrashort optical pulses form on the value of applied magnetic field for the case of square lattice is represented in Fig. 3.

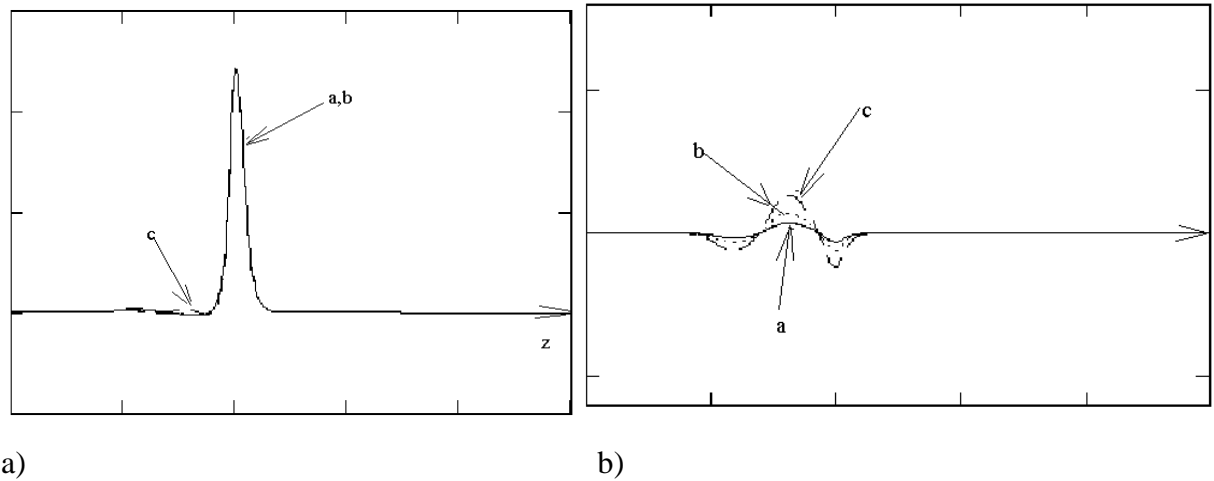

a)                                              b)

Fig. 3. Electric field determined by the equations (2, 6, 7) depending on constant magnetic field. Square lattice. The non-dimensionalized coordinate is plotted along the *x*-axis, the non-dimensionalized value of electric field (the unit is equal to $10^7$ V/m) is plotted along the *y*-axis. Fig. a) the primary polarization field, fig. b) the orthogonal polarization field. The line b) indicates the magnetic field is 2 times greater than the line a), the line c) indicates the magnetic field is 4 times greater than the line a). v/c=0,95.

It should be noted that the maximum value of electric-field intensity of orthogonal polarization pulse is proportional to the value of constant magnetic field. That is in good agreement with the hazarded conjecture and included in (7). Moreover, a change in the primary polarization pulse shape is observed with increase in the constant magnetic field. It can be directly attributed to the orthogonal polarization pulse field reaction on the current carriers providing the current which leads to initiation of the primary polarization field. By symmetry of the problem this is caused by phenomena similar to Hall effect, that is the deflection of moving under electromagnetic field pulse electrons by an external magnetic field.

Dependence of ultrashort optical pulses form on the value of applied magnetic field for the case of triangular lattice is represented in Fig. 4.

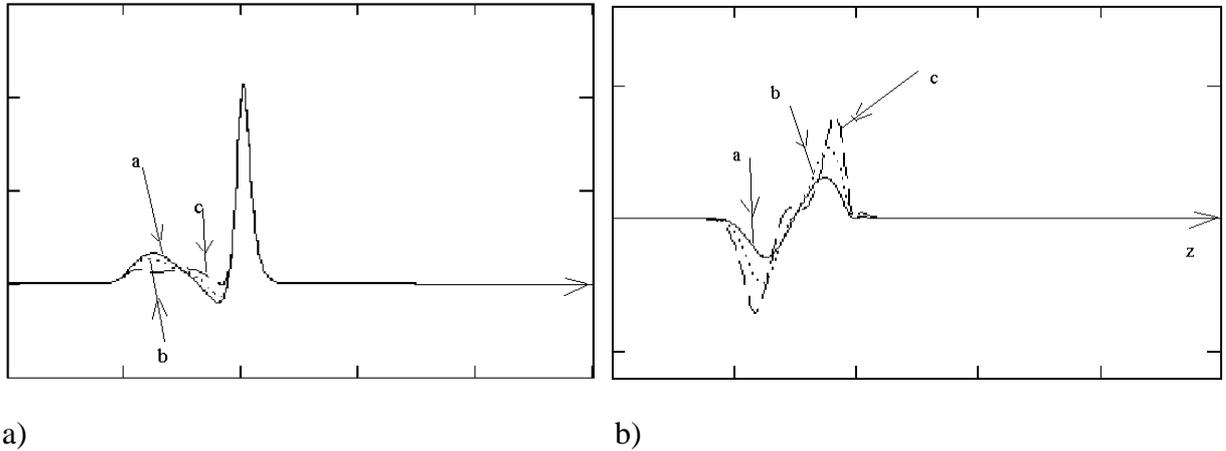

a)  b)

Fig. 4. Electric field determined by the equations (2, 6, 7) depending on constant magnetic field. Triangular lattice. The non-dimensionalized coordinate is plotted along the *x*-axis, the non-dimensionalized value of electric field (the unit is equal to $10^7$ V/m) is plotted along the *y*-axis. Fig. a) the primary polarization field, fig. b) the orthogonal polarization field. The line b) indicates the magnetic field is 2 times greater than the line a), the line c) indicates the magnetic field is 4 times greater than the line a). v/c=0,95.

It should be noted that in this case the constant magnetic field exerts antihunting effect on the ultrashort pulse of primary polarization and in stronger field this pulse less changes its shape. This can be useful in devices of optical signal processing. It should be also noted that like earlier the maximum value of electric-field intensity of orthogonal polarization pulse is proportional to the value of constant magnetic field. The pulse of orthogonal polarization for the triangle quantum dots lattice is generated more effectively than in the case of the square lattice.

The results of numerical calculations have demonstrated that dependence of ultrashort optical pulse form on parameter characterized the spectrum with overrunning of the nearest neighbors taken into account for the square lattice has the form as in Fig. 5.

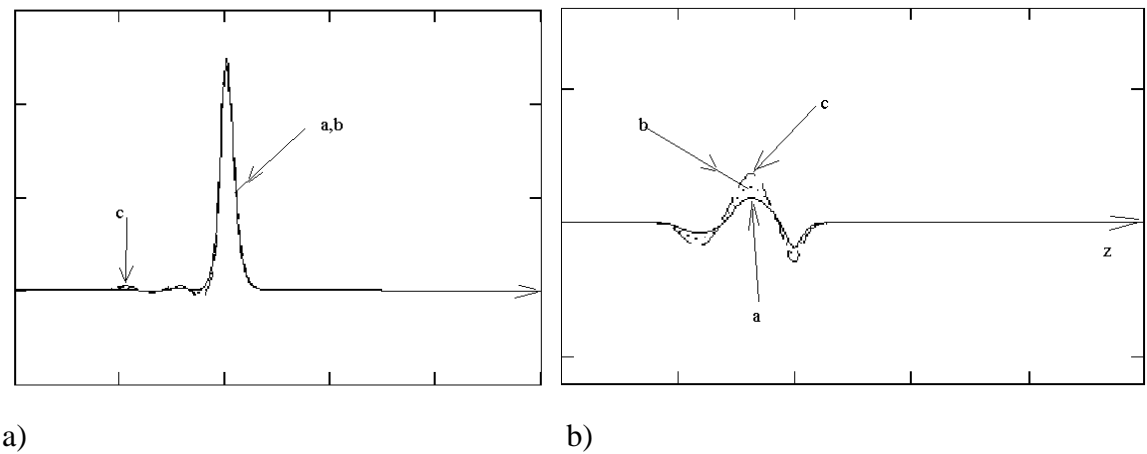

a)  b)

Fig. 5. Electric field determined by the equations (2, 6, 7) depending on parameter $\sigma$ of the spectrum. Triangular lattice. The non-dimensionalized coordinate (the unit is equal to $3 \cdot 10^{-4}$ m) is plotted along the *x*-axis, the non-dimensionalized value of electric field (the unit is equal to $10^7$ V/m) is plotted along the *y*-axis. Fig. a) the primary polarization field, fig. b) the orthogonal polarization field. The line b) indicates the parameter $\sigma$ is 2 times greater than the line a), the line c) indicates the parameter $\sigma$ is 3 times greater than the line a). v/c=0,95.

Such dependences can be explained with the circumstance taken into account that the term with $\sigma$ makes the equation (2) integrable even in the case of zero magnetic field. It means that the formation of ultrashort pulse tail is taking place. An increase in the maximum value of electric-field intensity of orthogonal polarization pulse with an increase in the parameter $\sigma$ can be explained by an increase in charge carrier mobility. This in its turn involves more effective deflection of the charge carrier by the constant magnetic field and thus an increase in corresponding current which initiates a pulse of a different polarization.

**4. Conclusion**

The results of the performed investigation allow us to make the following main conclusions:

1. The equation describing the dynamics of electromagnetic field in a quantum dot system with applied constant magnetic field was obtained.
2. The obtained efficient equation was numerically analyzed. A stable propagation of ultrashort optical pulses with non-zero area and a pulse formation with orthogonal polarization are revealed.
3. Dynamics and pulse form are highly dependent on the lattice geometry and in the case of triangular lattice stabilization of the pulse form by constant magnetic field is possible.
4. An increase in constant magnetic field value implies both pulse generation with orthogonal polarization and the change in the form of pulse with primary polarization.
5. The constant magnetic field value is the main cause for the pulse formation with orthogonal polarization and the change in that value allows varying the pulse form and amplitude.

**References**

1. Bass F.G., Bulgakov A.A., Tetervov A.P. High frequency properties of semiconductors with superlattices. M.: Nauka, 1989 – 288p.
2. Herman M. Semiconductor superlattices. M.: Mir, 1989 – 207p.
3. Valiev K.A., Kokin A.A. Quantum computers: hope and reality. "Regular and chaotic dynamics". Izhevsk, 2001.

4. T. Tanamoto, Phys.Rev. A, 61, 022305 (2000).

5. Ignatov A.A., Romanov J.A. Self-induced transparency in semiconductor superlattices//FTT. - 1975. - v.17. - N11. - C.3388-3389.

6. E.M. Epshtein. Solitons in superlattices//FTT. - 1977. - v.19. - N11. - C.3456-3458.

7. Landau L.D., Lifshitz E.M. Theoretical physics. T. II. Field theory. Moscow. Nauka, 1988. 512 p.

8. M.B. Belonenko, E.V. Demushkina, N.G. Lebedev. Journal of Russian Laser Research **27**, 457 (2006).

9. Romanov Yu A and Romanova J Yu 2004 *Solid State Phys.* **46** 164.

10. Polynovskij V.M. Nonlinear conductivity of semiconductor superlattices in strong magnetic field//FTP.- 1980. v.14.-N6.-p. 1215-1217.

11. E.M. Epshtein. Radiophysics, v.22, p. 373, 1979.

12. Bonch-Bruevich V.L., Kalashnikov S.G. Semiconductors physics. M. Nauka, 1990,865 p.

13. Lebwohl P A and Tsu R 1970 *J. Appl. Phys.* **41** 2664.

14. *N.S.Bakhvalov, Numerical Methods: Analysis, Algebra and Ordinary Differential Equations (Nauka, Moscow, 1975; Mir, Moscow, 1977).*

15. *Young-Woo Son, Marvin L. Cohen, Steven G. Louie.* Energy Gaps in grapheme nanoribbons // Physical Review Letters – 2006 – V. 97, N. 21. – P. 216803.

16. Kruchkov S.V., Fedorov E.G. Stabilization of solitons form in superlattices with spectrum overrunning of the nearest neighbors in field of nonlinear wave//FTP.- 2002. v.36.-N3.-p. 326-329.